\documentclass[sigconf]{acmart}
\AtBeginDocument{%
  \providecommand\BibTeX{{%
    \normalfont B\kern-0.5em{\scshape i\kern-0.25em b}\kern-0.8em\TeX}}}

\setcopyright{acmcopyright}
\copyrightyear{2018}
\acmYear{2018}
\acmDOI{XXXXXXX.XXXXXXX}

\acmConference[CVMP 2023]{the 20th ACM SIGGRAPH European Conference on
Visual Media Production}{Dec.\ 1--2}{London, UK}
\acmYear{2023}
\copyrightyear{2023}
\setcopyright{none}

\usepackage{graphicx}
\begin{document}

\title{DECORAIT - DECentralized Opt-in/out Registry for AI Training}

\author{Kar Balan, Alex Black, Andrew Gilbert}
\email{{ k.balan | a.black | a.gilbert } @ surrey.ac.uk}
\orcid{1234-5678-9012}
\affiliation{%
  \institution{University of Surrey}
  \city{Guildford}
  \country{UK}
}

\author{Simon Jenni, Andy Parsons, John Collomosse}
\email{{jenni | andyp | collomos}@adobe.com}
\affiliation{%
  \institution{Adobe Inc.}
  \streetaddress{....}
  \city{San Jose}
  \country{US}
}
\makeatletter
\DeclareRobustCommand\onedot{\futurelet\@let@token\@onedot}
\def\@onedot{\ifx\@let@token.\else.\null\fi\xspace}
\def\eg{\emph{e.g}\onedot} \def\Eg{\emph{E.g}\onedot}
\def\ie{\emph{i.e}\onedot} \def\Ie{\emph{I.e}\onedot}
\def\cf{\emph{c.f}\onedot} \def\Cf{\emph{C.f}\onedot}
\def\etc{\emph{etc}\onedot} \def\vs{\emph{vs}\onedot}
\def\wrt{w.r.t\onedot} \def\dof{d.o.f\onedot}
\def\etal{\emph{et al}\onedot}
\makeatother

\begin{abstract}
We present DECORAIT; a decentralized registry through which content creators may assert their right to opt in or out of AI training as well as receive reward for their contributions.  Generative AI (GenAI) enables images to be synthesized using AI models trained on vast amounts of data scraped from public sources. Model and content creators who may wish to share their work openly without sanctioning its use for training are thus presented with a data governance challenge. Further, establishing the provenance of GenAI training data is important to creatives to ensure fair recognition and reward for their such use.  We report a prototype of DECORAIT, which explores hierarchical clustering and a combination of on/off-chain storage to create a scalable decentralized registry to trace the provenance of GenAI training data in order to determine training consent and reward creatives who contribute that data. DECORAIT combines distributed ledger technology (DLT) with visual fingerprinting, leveraging the emerging C2PA (Coalition for Content Provenance and Authenticity) standard to create a secure, open registry through which creatives may express consent and data ownership for GenAI.
\end{abstract}



\begin{teaserfigure}
  \includegraphics[width=\textwidth]{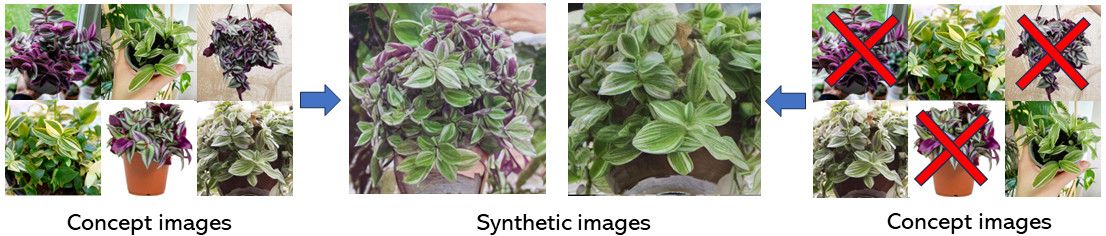}
  \caption{DECORAIT enables creatives to register consent (or not) for Generative AI training using their content, as well as to receive recognition and reward for that use. Provenance is traced via visual matching, and consent and ownership registered using a distributed ledger (blockchain). Here, a synthetic image is generated via the Dreambooth\cite{dreambooth} method using prompt "a photo of \textit{[Subject]}" and concept images (left). The red cross indicates images whose creatives have opted out of AI training via DECORAIT, which when taken into account leads to a significant visual change (right). DECORAIT also determines credit apportionment across the opted-in images and pays a proportionate reward to creators via crypto-currency micropyament.}
  \label{fig:teaser}
\end{teaserfigure}


\maketitle

\section{Introduction}
Generative AI (GenAI) models such as ChatGPT and Stable Diffusion \cite{chatgpt,sd}  are transforming creative workflows through their ability to synthesize content given only high-level direction.  GenAI models are typically trained by sampling millions of media items harvested from public data sources. Yet this practice has raised concerns over data governance, specifically over creatives’ agency to opt in or out of the use of their work for GenAI training.  This has led to several legal challenges to the creators of GenAI models, stemming from the concerns over potential rights infringement -- particularly of digital images.  Furthermore, creatives are not currently enabled to receive recognition or reward for their contribution to GenAI images through the use of their content in training.

We envision a future creative economy for content delivery services, such as stock photography platforms, which enables the commercialization of creative content and the contribution of it to ethically and consensually built datasets for GenAI training. These platforms are responsible for enabling their users, contributors, and collaborators to express consent over their data being used in GenAI training in a secure, persistent, and interoperable way. Such capability is grounded in strong provenance signals for GenAI training data, that enable creatives to register ownership and means for payment for GenAI use as well as their consent for that use.

To this end, we propose DECORAIT, a decentralized registry for GenAI training that enables creators to express consent, or otherwise, for their images to be used in AI training, as well as enabling them to receive reward when such use occurs.  Our work follows emerging community trends toward centralized, commercial opt-out services.  For example, {\em spawning.ai} maintains lists of opted-out URL patterns (from individual links to entire domains).  GenAI models can match against these lists to exclude content from training. However, a URL list may not capture all instances of a creator's content online.  Moreover, scaling up multiple individually managed databases to track opt-out raises data consistency and interoperability challenges. The protocol of the future creative economy also ought to ensure the contributing creatives to GenAI can be recognized and rewarded for their creative assets when their particular content or style is identified to have contributed to specific synthetic media. DECORAIT addresses these issues through three contributions:

\begin{enumerate}
    \item{We propose a {\bf fingerprint-based content similarity score}, followed by a {\bf credit apportionment scheme} to match images and reward creatives for their training content most correlated with generated synthetic media.}

    \item{A {\bf sharded decentralized search index} using distributed ledger technology (DLT), in which a content fingerprint distilled from the image provides a key to register and robustly query opt-in/out information.  We propose a hierarchical approach to scale vector search of this index and a hybrid on/off-chain approach to query processing.}
    
    \item{We leverage the emerging {\bf Coalition for Content Provenance and Authenticity (C2PA)} standard to express consent and payment preferences via cryptographically signed asset `manifests'. These manifests are stored within a distributed file system (IPFS) and referenced by hashed URL link via the DECORAIT DLT search index.}

\end{enumerate}

Without loss of generality, we demonstrate DECORAIT within the pipeline of Dreambooth \cite{dreambooth} which enables specialization of diffusion models to generate novel renditions of a specific subject provided via exemplar training images.  Dreambooth provides a suitable use case as it enables GenAI model users to assure the assets they intend to leverage for model personalization have been opted-in for AI training. Additionally, the proposed system enables the fair recognition and reward of those contributing creatives. 
 We could imagine a future for stock photography in which contributors receive payments not only through direct licensing (as now), but automatically via DECORAIT's ability to provide downstream recognition and persistent crediting of the contributing creators to GenAI. Fair monetary reward is encouraged via our apportioning algorithm, coupled with the transparency and auditability of crypto-currency payments processed using DLT.

\section{Related Work}

{\bf Distributed Ledger Technology (DLT)}, colloquially 'blockchain', ensures the immutability of data distributed across many parties without requiring those parties to trust one another or any central authority \cite{Narayanan2016}. While the original and dominant use is cryptocurrency tokens (\eg Bitcoin \cite{bitcoin}), emerging use cases include digital preservation  \cite{Lemieux2016}, supply chain and media provenance \cite{Walport2015,Holmes2018}.  DLT has been used to track ownership of media via the ERC-721 Non-Fungible Token (NFT) standard \cite{nftsurvey}, although NFT lacks a rights or permissions framework \cite{nftlaw}. Recently, Ekila explored tokenized rights in NFT \cite{ekila}.  DLT was analyzed for media integrity in ARCHANGEL \cite{archangel}; digital records were hashed and used to tamper-proof archival records.  Perceptual hashing extended ARCHANGEL from documents to images \cite{arch2} and videos \cite{arch3}.  Our work uses perceptual hashes for search; as a key to resolving image fingerprints to data on training consent. Recent advances in proof of stake and Layer 2 solutions scale DLT for improved throughput and reduced climate impact, yet scalable storage remains challenging. Peer-to-peer (p2p) distributed file-sharing technologies such as the Interplanetary File System (IPFS \cite{benet2014ipfs}) are used to address this.

\noindent {\bf C2PA} is an emerging metadata standard for embedding content provenance information (`manifests') in media files (`assets') \cite{c2pa}.  Manifests are signed via public-key pair and describe facts about asset provenance, such as who made it, how and using which ingredient assets. These facts are called `assertions'. C2PA initially focused on trusted media \cite{cai} and journalism \cite{origin} use cases. Recently, C2PA (v1.3) described a \texttt{training-mining} assertion in which creators may set flags to opt in or out of GenAI training, which we leverage in our work.   Unfortunately, C2PA metadata is stripped by non-compliant platforms (\eg social media) or attackers. Therefore, we use perceptual hashing to match content to manifests.

\noindent {\bf Content Fingerprinting} identifies content robustly in the presence of degradation or rendition (format, quality, or resolution change) and minor manipulation. Perceptual hashing \cite{nguyen2021,Black_2021_CVPR,Bharati2021} and watermarking \cite{devi2009,rosteals} have been used to match content. Fingerprinting has also been used to detect and attribute images to the GenAI models that made them \cite{yu2021responsible}.

\noindent{\bf Diffusion models} lay at the foundation of most recent advances in GenAI \cite{ho2020denoising, sdxl, ldm,imagen,dalle2}. Such models are commonly trained on millions or even billions of images to gain the ability to synthesize diverse and high-quality images consistently. Diffusion models have shown substantially superior performance in comparison to GANs \cite{dhariwal2021diffusion}. However, they have also been shown to memorize content and style from training data to a higher degree than GANs \cite{somepalli2022diffusion}, phenomenon attributed to the presence of duplicated image data \cite{carlini2023extracting, somepalli2022diffusion} within the training data. \cite{somepalli2023understanding} showed that content and style memorization is an even greater concern, specifically in text-conditioned diffusion models, due to duplicated captions within the training data, with data replication not commonly occurring in unconditional diffusion models. This further accentuates the need to involve creatives and obtain consent to use their creative content in the GenAI training pipeline. The present work lays the groundwork for such a system, querying and registering the creatives' opt-in or out decision on GenAI training, as well as offering a pipeline to reward creatives for using those assets in GenAI.

\noindent{\bf Model personalization} methods are techniques which enable diffusion models to be customized to synthesize novel renditions of a specific subject in different contexts. Recently, both training-free adaptation \cite{shi2023instantbooth} and fine-tuning \cite{kumari2023multiconcept, dreambooth} have been explored to perform customization to object instances. 
 In this work, we utilize the Dreambooth \cite{dreambooth} technique for model personalization, which fine-tunes a pre-trained text-to-image diffusion model - the base model - using a small set of `concept' images depicting a specific subject. The subject is thus embedded in the output domain of the model which learns to bind it to a unique identifier (token), which can then be used as part of the prompt to synthesize the subject in new and diverse contexts. We use Dreambooth to demonstrate the DECORAIT system, aiding in the training pipeline by identifying opted-in assets from a stock photography website available to train a personalized instance of a diffusion model.

\section {Tracing and Describing Image Provenance}

We begin by describing how images are matched to trace visual provenance. We use this approach to `fingerprint' training images in order to robustly match to entries in the DECORAIT registry, thereby accessing data on consent status and creator wallet addresses which are encoded via the C2PA open standard (subsec. \ref{sec:c2pa}).  A second pair-wise model enables both verification of such matches and correlation between synthetic and training data for credit apportionment.

\subsection{Fingerprinting and match verification}

In order to reliably match training images at scale, we employ two modules.  First, a contrastively trained model to extract compact embeddings for measuring image similarity, and second a model which classifies whether the closest matching images in that embedding are true matches. The latter is motivated by the difficulty of thresholding image similarity distances at scale whilst retaining practical accuracy levels.  The classifier probability serves both as a match verification check and a score to drive credit apportionment.

\subsubsection{\bf{Fingerprinting Model}} We adapt the fingerprinting technique described in \cite{Black_2021_CVPR} to obtain compact embeddings of the images within the registry's corpus, allowing robust visual content attribution and search.
\label{ref:fprint}
The resulting fingerprint is a compact embedding (256-D) of a CNN, contrastively trained to be discriminative of image content whilst robust to image degradations and manipulations, to model content transformations common as images are shared online.  The smodel is trained through a contrastive learning objective \cite{chen2020simple}.
Let $\phi_i=E(x_i)\in \mathbb{R}^{256}$ be the feature vector obtained as the output of a ResNet-50 encoder for an image $x_i$ and $\hat{\phi}_i$ represent an embedding of a differently augmented version of $x_i$.
The training objective is given by
\begin{equation}
\mathcal{L}_{C} = - \sum_{i\in \mathcal{B}}   \log \left( \frac{d\left( \phi_i, \hat{\phi}_i \right)}{ d\left( \phi_i, \hat{\phi}_i \right) + \sum_{j \neq i \in \mathcal{B}} d\left( \phi_i, \phi_k \right)} \right),
\end{equation}
where $d(a, b):= \exp \left(\frac{1}{\lambda}  \frac{ {a}^\intercal { b}}{\Vert {a} \Vert_2 \Vert {b} \Vert_2} \right)$ measures the similarity between the feature vectors $a$ and $b$, and $\mathcal{B}$ is a large randomly sampled training mini-batch \cite{ekila}.

 In terms of data augmentation, in addition to the typical techniques used in contrastive learning such as colour jittering and random cropping, we consider minor manipulations, benign modifications and degradations of image content due to noise, format change and recompression, resolution change (resize), and several other degradation manipulations studied in \cite{hendrycks2019robustness}.  This is because images may be reshared online and be subject to many such transformations and renditions, and we wish to match regardless of these.

\subsubsection{\bf{Verification and Apportionment Model}}
\label{sec:apportion}
Provided a shortlist of the top-K candidate matches from the previous fingerprinting step, we verify image matches through an additional pair-wise comparison between the query image and each candidate match retrieved. The spatial feature maps derived from the fingerprinter model are used to compare the images as follows.

\begin{figure}[h]
    \centering
    \includegraphics[width=0.8\columnwidth]{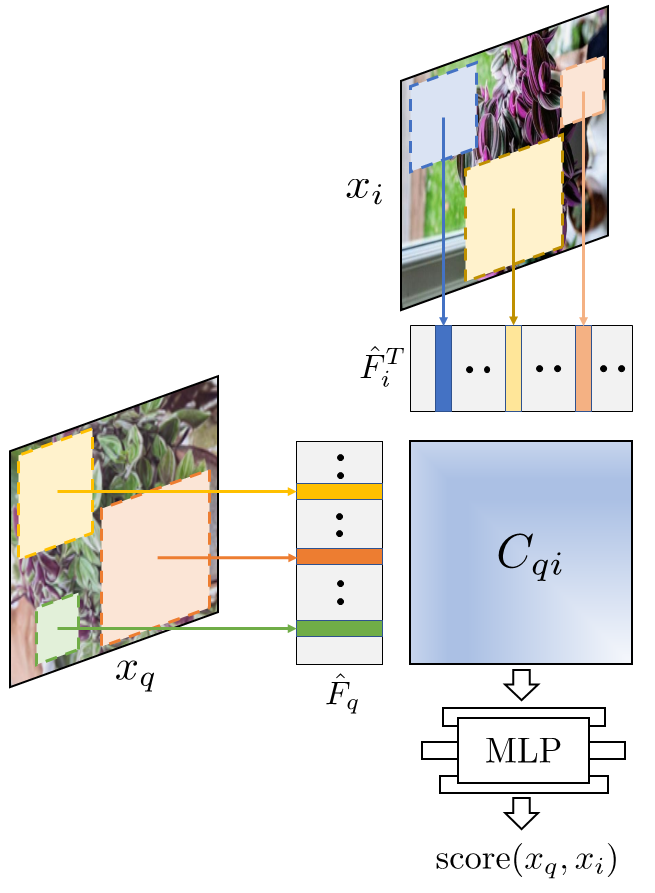}
    \caption{Match Verification Model. Two images are compared at multiple scales to robustly find (partial) matches. The model extracts multiple aggregated feature vectors from the two feature maps corresponding to numerous image patches of different sizes and positions. These features (collected in $\hat{F}_q$ and $\hat{F}_i$) are then used to compute the feature correlation matrix $C_{qi}$, which is fed to an MLP to compute a final score.}
    \label{fig:verifier}
\end{figure}

Let $F_q\in \mathbb{R}^{H\times W \times D}$ be the feature map for a query image $x_q$ and let $\{F_i\}_{i=1}^k$ be the $k$ corresponding retrieval feature maps.
Each feature map is processed with a $1\times1$ convolution to reduce the dimensionality to $\frac{D}{4}$ and then numerous pooled descriptors from a set of 2D feature map windows $\mathcal{W} \subset [1, H] \times [1, W]$ are extracted, similar to R-MAC \cite{tolias2015particular}.
Let $f^q_{w}\in \mathbb{R}^\frac{D}{4}$ denote the GeM-pooled \cite{tolias2015particular} and unit-normalized feature vector for a window $w\in \mathcal{W}$ and feature map $F_q$.
In contrast to \cite{tolias2015particular}, the window-pooled feature vectors are not averaged, but collected as:
\begin{equation}
    \hat{F}_q = [f^q_{w_1}, \ldots, f^q_{w_{|\mathcal{W}|}}] \in \mathbb{R}^{|\mathcal{W}| \times \frac{D}{4}},
\end{equation}
where $w_i \in \mathcal{W}$ and the number of windows is $|\mathcal{W}|=55$ in practice.
The feature correlation matrix is then computed as:
\begin{equation}
    C_{qi} = \hat{F}_q \hat{F}^T_i \in \mathbb{R}^{|\mathcal{W}|\times |\mathcal{W}|}.
\end{equation}
These feature correlations are then flattened and fed to a 3-layer $\operatorname{MLP}$, which outputs a similarity score between query $q$ and retrieval $i$.
To make the model symmetric w.r.t. its inputs, the match score between images $x_q$ and $x_i$ is defined as 
\begin{equation}
    \operatorname{apportion}(x_q, x_i) = \sigma \big( \operatorname{MLP}(C_{qi}) + \operatorname{MLP}(C_{iq}) \big),
    \label{eq:apportion}
\end{equation}
where $\sigma$ represents a sigmoid activation. The model is illustrated in Fig.~\ref{fig:verifier}. 

To train the model, positive example pairs are built via a strong data augmentation protocol, similar to the data augmentation step in the fingerprinter model training. This protocol includes colour jittering, blurring, random resize cropping, and random rotations. A hard negative mining approach is used to generate challenging negatives.

For the sampling of negatives, the global average-pooled feature maps of query and queued examples are compared via cosine similarity. Given pairs of true and false matches, the model is trained with a standard binary cross-entropy loss. During training, the backbone feature extractor from the fingerprinter model is frozen.

\subsection{Encoding consent and ownership}

\label{sec:c2pa}

The Coalition for Content Provenance and Authenticity (C2PA) standard aims to aid internet users' trust decisions about digital assets they might come across on platforms such as social media or news websites. Recent work also employs C2PA as a tool to encode provenance information within synthetically generated media, including within its metadata details about the model used to create it, as well as its training data \cite{ekila}.

A `manifest' is a data packet that may be bound to digital assets at creation time or post-factum. This manifest embeds facts about the provenance of a digital asset within its metadata. These facts are referred to as `assertions'. They may include information such as who created the asset, how it was made, what hardware and software solutions aided in its creation, and any edits it may have undergone since its creation.  This data is cryptographically signed to prevent tampering. Signing C2PA manifests requires that the signer uses their private key and public certificates, following the Public Key Infrastructure (PKI). This assures that the consumer makes trust decisions about the asset based on the identity of the manifest signer. A certification authority (CA) conducts real-world verification to ensure signing credentials are only issued to trusted, non-malicious actors \cite{c2pa}.

Additionally, C2PA manifests may bear information about other "ingredient" assets used in the creation process. These ingredients may point at assets, each bearing its own C2PA manifest describing its provenance. As such, C2PA encodes a graph structure with the root at the current asset and branching out to its ingredient assets.   The C2PA standard describes this ingredient model in terms of creation of classical images (and other media assets) but we use it in DECORAIT to describe how Dreambooth models may be created from their training concept images, and how synthetic images are created from their Dreambooth model as an ingredient. 

Recently, C2PA (v1.3) introduced several training-mining assertions in which creators may set flags to opt in or out of GenAI training within manifests. These flags are \emph{data\_mining, ai\_inference, ai\_generative\_training} and \emph{ai\_training}. We leverage these flags to encode consent in DECORAIT.

C2PA manifests also support the inclusion of DLT-based wallet addresses.  For example, in Adobe Photoshop, any DLT wallet address linked to a user's Adobe identity may be recorded in the C2PA metadata of an exported image. In the following sections, we show how this wallet information, embedded immutably within assets at creation-time, may be leveraged to reward creatives when their images are used to train GenAI.

\section{DECORAIT System Architecture}

\begin{figure*}[h]
    \centering
    \includegraphics[width=1\textwidth]{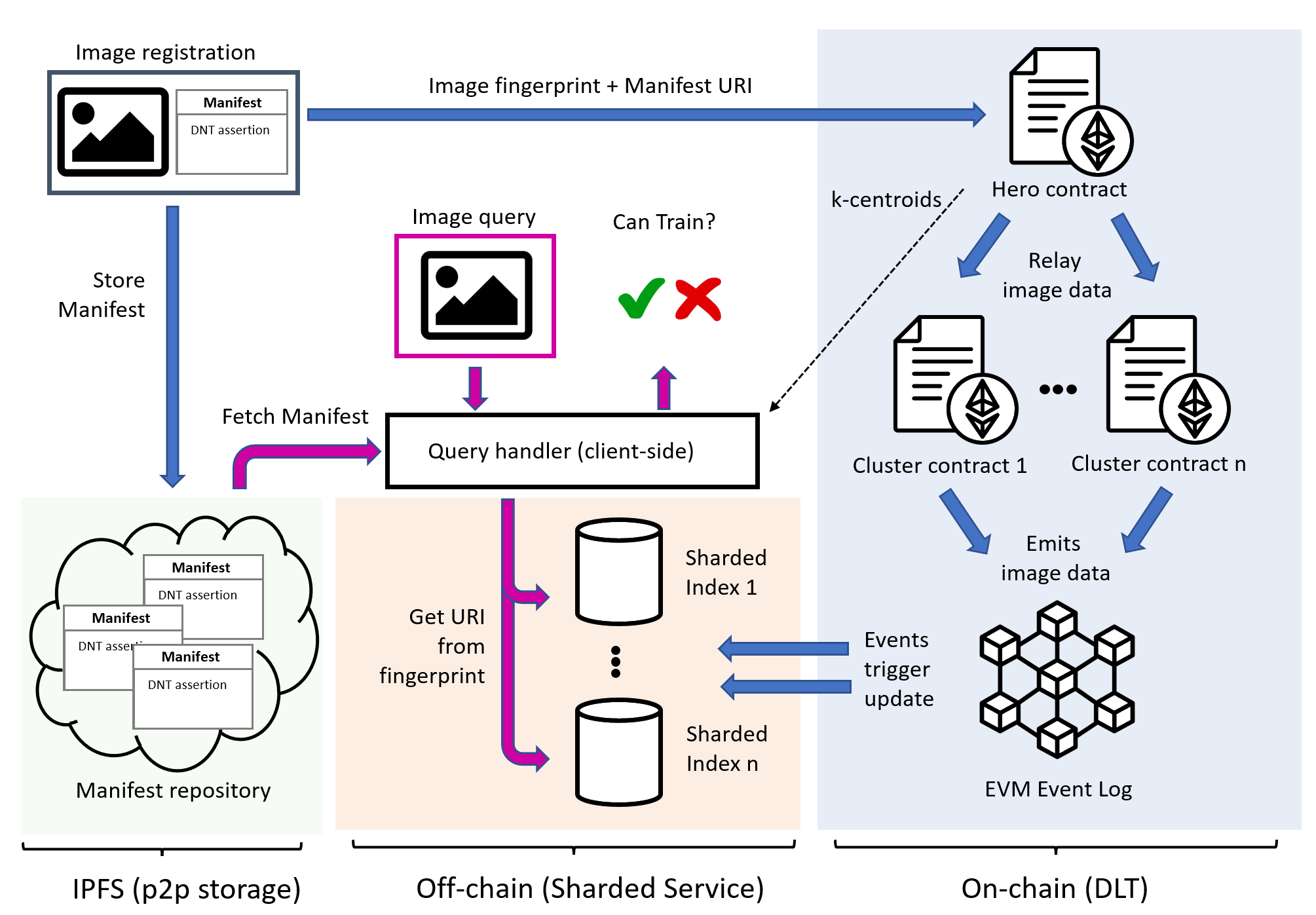}
    \caption{Overview of DECORAIT.  1) Ingest (blue): An image is fingerprinted by the client, and the hash is passed to the Hero contract, which determines on-chain which of the sharded (cluster) contracts will handle the ingest.  The cluster contract emits an event recording the fingerprint (key) and IPFS URI (value) of the C2PA manifest, which the client stores on IPFS.  The relevant off-chain sharded index listens for on-chain updates from its respective contract.  2) Query (pink): An image is fingerprinted by the client, and k-centroid data is used to determine which index shard to query with the fingerprint (key) to obtain the C2PA manifest URI (value).  The client decides on whether GenAI training is permitted using the manifest.  The diagram reflects the recommended variant (E-FOF) of DECORAIT (\cf Table~\ref{tab:variants}).}
    \label{fig:arch}
\end{figure*}

\label{archit}

    DECORAIT is a decentralized search index, performing {\em key-value} lookups using a robust image fingerprint (subsec.~\ref{ref:fprint}) as the {\em key}.  The {\em value} is a URI, resolvable to a C2PA manifest indicating permission to train.  A scalable solution demands: 1) persistent distributed storage of manifests; 2) a distributed and immutable lookup operated via an open model without recourse to a centralized trust.
Fig.~\ref{fig:arch} provides an overview of DECORAIT, which addresses these decoupled requirements by: 1) storing manifests on IPFS, where URIs are formed using a CID -- a bit-wise (SHA-256 \cite{sha}) content hash; 2) using a hybrid on/off-chain solution to create a sharded search index (subsec. \ref{sec:searchindex}).  In Sec. \ref{sec:eval}, we explore empirical trade-offs in defining the boundary between on and off-chain computation for the search and the optimal level of sharding.

\subsection{Decentralized Fingerprint Index}
\label{sec:searchindex}

All images within DECORAIT undergo visual fingerprinting using the approach outlined in Sec. \ref{ref:fprint} to enable large-scale retrieval of visually similar assets upon querying the registry. We adopt a hierarchical approach to share the search index, applying k-means clustering to fingerprints computed from a representative (1M) image sample. The resulting k-centroids subdivide the fingerprint hash space into $k$ shards. Recursive sharding is possible, but experiments focus on a single level. We shard the index using $k+1$ DLT smart contracts deployed on a local Ethereum test-net; one contract per each of the $k$ clusters, plus a single entry point -- the `hero' contract -- to orchestrate the sharding.  The hero contract performs the k-NN assignment of fingerprints to the k-centroids, delegating operations (\eg ingest, query) to be handled by the smart contract of the closest cluster (and so, shard).  

The contracts are implemented in Solidity, which does not support floating point math. We convert the 256-D floating point fingerprinting embeddings into integers as fixed-point (10\textsuperscript{15} precision), a workaround for applying ML operations on DLT \cite{microsoftpaper}. 

\subsection{Hybrid on/off-chain variants}

We explore several design choices for implementing our system, evaluating three main variants (Table \ref{tab:variants}).  In particular, we explore options for persisting the key-value store (here used to map image fingerprints to manifest URIs) and on/off-chain options for implementing the shard assignment and retrieval processes.

\subsubsection{\bf{Image Fingerprints and Data Storage}}

DLT storage patterns commonly persist data in two main ways: 1) {\em in-contract} \ie within the state of a smart contract (as with NFTs), or 2) on the {\em event log}, a ledger of signals/exceptions emitted from smart contract code (as with cryptocurrency transactions).  In our experiments, we use mnemonics prefixed E- to indicate variants using the event log and C- to indicate variants using in-contract storage.

Shards are described by k-centroid data from clustering in fixed point (256 integers) form.  Fingerprints are similarly represented. These 256-D data are stored as strings in the event log but may be stored in integer arrays for in-contract storage. There is cost efficiency in storing strings over integer arrays.  However, there is a time cost in converting the strings for fixed point operation during ingest and query.  The transaction cost implications are quantified in subsec.~\ref{sec:costs}.

\begin{table}[h]
  \caption{Configuration of the three implementation variants.}
  \label{tab:variants}
\resizebox{\columnwidth}{!}{%
\begin{tabular}{cccc}
\hline
Action & C-OOO & E-OOF & E-FOF \\ \hline
Key-value storage & in contract & event log & event log \\ 
Shard centroid prediction on query & on-chain & on-chain & off-chain \\ 
Shard prediction on ingest & on-chain & on-chain & on-chain \\ 
Retrieval with-in shard & on-chain & off-chain & off-chain \\ \hline
\end{tabular}%
}
\end{table}


\subsubsection{\bf{Shard assignment and retrieval}}

To store (ingest) or retrieve (query) a key-value pair, it is necessary to match the fingerprint to its shard via a k-NN assignment operation against the k-centroids obtained during initial clustering.  In the case of queries, the retrieval is performed by matching against each key (fingerprint) stored within the key-value store of that shard.  In Table~\ref{tab:variants}, we use 'O' to indicate on-chain and 'F' to indicate off-chain computation for each matching operation and compare the efficiency of these variants in Sec.~\ref{sec:eval}.

\subsubsection{\bf{Smart contract interaction}}

The hero contract receives all operations and transactions in all variants.  When ingesting a fingerprint to the registry, the hero contract reads it and calls the respective shard contract, which stores the key-value pair within its own contract (C-) or the event log (E-). In all cases, the smart contract performs the sharding via on-chain operations, which safeguards the integrity of the shards against inaccurate or malicious additions that could otherwise "infect" the clusters.


When querying a fingerprint, the on-chain variant (C-OOO) proceeds similarly - determining the shard and delegating the retrieval process to the relevant smart contract.  The retrieval is performed on-chain in this case. In variants E-OOF and E-FOF, the query processing is partly delegated to off-chain processes.  A web service is provided for each shard, which listens to the event log emitted by the smart contract of its respective shard.  When submitting a query, the hero smart contract determines the appropriate shard index to call.  This may be done on-chain (per the ingestion flow) or off-chain using k-centroid data from the hero contract.  The relevant shard's web service performs the retrieval in both cases.  Using off-chain processing mitigates computational costs as the index scales, as we now show.  Fig~\ref{fig:arch} shows the interaction of the web service and smart contracts.

\subsection{DECORAIT in the GenAI Workflow}
\begin{figure*}[h]
    \centering
    \includegraphics[width=1\textwidth]{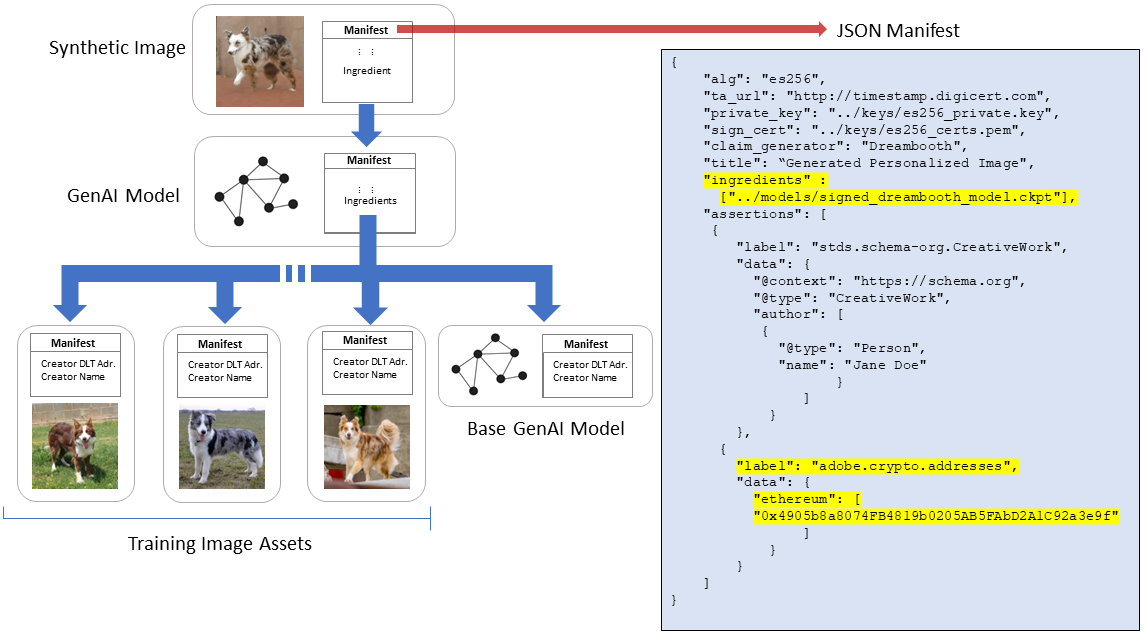}
    \caption{Provenance graph of synthetic media, as may be encoded via C2PA manifests. Left: starting from the generated image, the specialized Dreambooth model is listed as ingredient in its C2PA manifest. In turn, the Dreambooth model links to the specialization set of images and the base text-to-image Stable Diffusion model, which then may list as ingredient an archive of its entire training image corpus. Right: example JSON C2PA manifest accompanying a synthetic image, with highlighted ingredient and DLT wallet address assertions (the latter using the schema of a commercial image editor).}
    \label{fig:provenance}
\end{figure*}

We now describe how DECORAIT integrates with the GenAI training process to determine consent and how subsequently generated synthetic images may be traced to pay a reward to the creators who contributed that training data.

\subsubsection{\bf{Training Consent}}

To ensure the creatives who authored the images have consented to their assets being used for training, each image is queried against the DECORAIT registry.  As described in subsec.~\ref{sec:searchindex} the fingerprint embedding is used to identify the closest visual matches within the decentralized search index.  These are verified using the apportionment model (subsec.~\ref{sec:apportion}) to obtain the closest match and so, a decision on training consent for each of the images. As described in subsec.~\ref{sec:c2pa}, this information is embedded within the C2PA manifest accompanying each image on the DECORAIT system. We envision a future in which stock photography sites might parse and show this consent information by default, enabling users to select only the opted-in images when sourcing data for GenAI training.

\subsubsection{\bf{Encoding Synthetic Image Provenance}}

We further leverage the C2PA standard to encode the provenance of the newly generated synthetic image, cryptographically tying it to the "ingredient"  set of concept images and GenAI model. Thus, using the C2PA manifest of the generated image it is possible to trace both the model that generated it (the personalized model, and in turn, its base model), as well as the data used to personalize it. Specifically, the C2PA "ingredient" assertion is used to indicate the image dataset as ingredients to the fine-tuned model, as well as the base model. The personalized, fine-tuned model is then listed as an ingredient within the manifests of any synthetic images it generates. Thus, the synthetic image is tied to its ingredient assets listed above. This offers a complete creation provenance chain, immutably signed at creation time. Although in the case of finetuning models, the images from the dataset are individually included in the manifest, C2PA allows for manifests to be defined over archives of image collections for larger datasets. Figure \ref{fig:provenance} visualizes this relationship.

\subsubsection{\bf{Apportionment and Payment}}

Given a synthetic image, DECORAIT enables credit to be assigned across training data images in order to recognize and reward contribution. The set of training image ingredients is first identified by traversing the image's provenance graph, rooted in the manifest of the synthetic image. Similar to the training stage, DECORAIT again uses visual fingerprinting to perform matching within the decentralized search index to lookup the C2PA manifest of each training image --- including the DLT wallet address of each image's creator.

Credit is then assigned to each image proportional to a pair-wise score predicted by the apportionment model of subsec.~\ref{sec:apportion}: given a synthetically generated image $X_q \in \mathbb{R}^{H\times W \times 3}$, the  visual similarity of each training image $X_i$ 
in the identified concept set is scored via eq.~\ref{eq:apportion} yielding a weighting $w_{i}$:
\begin{equation}
    w_{i} = \max \big(\operatorname{apportion}(X_q, X_i)-\lambda, 0 \big),
\end{equation}
where $\lambda=0.7$ is an empirically set threshold for the visual similarity.  We then assign credit per image by normalizing these weights over all top-image matches in the concept set.  

Once the credit apportionment has been determined, payments may be processed securely and transparently over DLT. The authors' wallet addresses are extracted from the C2PA manifest associated with each image (Figure \ref{fig:provenance}, left).

In Sec.\ref{eval_dreambooth} we demonstrate how the DECORAIT system can be applied to a Dreambooth training pipeline, querying the registry for training consent, computing the similarity score, and apportioning credit amongst the set of concept images for any given generation.

\section{Evaluation}
\label{sec:eval}
We evaluate the relative performance and scalability of the three variants of the DECORAIT decentralized search index: C-OOO, E-OOF, E-FOF (\cf Table \ref{tab:variants}) concluding on the most performant variant. We then demonstrate the proposed variant of the DECORAIT system as applied to the use case of a Dreambooth training pipeline and demonstrate querying the registry, resolving to a decision on training consent of the images within the set of concept images, followed by processing payments using DLT based on our apportionment algorithm.

\subsection{Experimental Setup}

We evaluate using the LAION400M dataset \cite{laion400m}, comprising image-text pairs crawled from publicly available web pages. LAION400M is extensively used to train GenAI models. For our experiments, we sample a training corpus of 1M images and sign these with C2PA manifests setting the \texttt{ai\_generative\_training}, \texttt{data\_mining}, and \texttt{ai\_training flags} to `not allowed' to signify that the author has opted out of those images being used to train GenAI models.

The evaluation uses up to 1000 query images randomly sampled from the corpus, to which random augmentations are applied. The data augmentation process follows \cite{Black_2021_CVPR}. It aims to mimic the perturbations an image may suffer from repeated use, upload, download, and compression on the internet (\eg noise, and changes in resolution, quality, and format). In addition, we form a second query set of 100 unperturbed images. Lastly, we demonstrate the proposed DECORAIT system variant (E-FOF) within a Dreambooth model specialization pipeline.

\subsection{Evaluating Accuracy vs Sharding}

We evaluate the lookup's accuracy as a function of sharding (cluster count $k$) while maintaining a constant corpus size. 
 The accuracy is agnostic to the on/off chain implementation of storage and query lookup, but the performance (query speed) varies significantly. Results are reported in table \ref{tab:exp1} for 1000 queries. 

 There is a trend to slightly reduced accuracy as sharding ($k$) increases due to the risk of heavy perturbations mis-assigning image fingerprints to adjacent shards.  When no perturbation is present, the system performs with $100\%$ accuracy for all shard counts.  Yet, increasing sharding will reduce retrieval time (see below).  On this basis, we select $k=25$ as an appropriate sharding trade-off for the remainder of our experiments.

\begin{table}[h]
  \caption{Evaluating accuracy vs. shard count ($k$)  for a 0.5M corpus size.  The performance of on-chain shard prediction and within shard retrieval is studied for E-OOF. Shard prediction time increases, but retrieval time decreases as $k$ increases.}
  \label{tab:exp1}
\resizebox{\columnwidth}{!}{%
\begin{tabular}{cccc}
\hline
\begin{tabular}[c]{@{}c@{}}Clusters \\ (k)\end{tabular} & \begin{tabular}[c]{@{}c@{}}Accuracy\\  (\%)\end{tabular} & \begin{tabular}[c]{@{}c@{}}Cluster Prediction \\ Time (on-chain)(s)\end{tabular} & \begin{tabular}[c]{@{}c@{}}Retrieval Time \\ (off-chain) (ms)\end{tabular} \\ \hline
1(Baseline) & 92.3 & - & 46.157 \\ 
15 & 89.1 & 6.5 & 3.612 \\ 
25 & 89.1 & 10.3 & 2.461 \\ 
50 & 88.1 & 20.6 & 1.306 \\ 
100 & 87.3 & 35.2 & 0.721 \\ 
200 & 86.5 & 59.9 & 0.413 \\ 
500 & 86 & 113 & 0.284 \\ 
750 & 87.4 & 181.8 & 0.271 \\ 
1000 & 86.4 & 272.9 & 0.249 \\ \hline
\end{tabular}%
}
\end{table}

\subsection{Evaluating Performance vs Sharding}

Retrieval speed varies significantly for each of our three variants and comprises two processes: closest centroid (shard) prediction and retrieval within the shard. We evaluate the speed of the nearest centroid prediction as a function of shard count ($k$). The number of distance computations (between the query embedding and each cluster centroid) scales linearly with $k$ (Table \ref{tab:exp1}), and this becomes prohibitive (several seconds) for high-frequency transactions (queries) at $k=25$, though acceptable for bulk ingestion.  This suggests that variant patterns x-Oxx are not scalable at query time.

\begin{table}[h]
  \caption{Evaluating in-contract storage (C-OOO) for accuracy and speed as corpus size increases. Shard count $k=25$. Accuracy is good, but speed is poor relative to event-log variants.}
  \label{tab:exp2a}
\resizebox{\columnwidth}{!}{%
\begin{tabular}{cccc}
\hline
Corpus & \begin{tabular}[c]{@{}c@{}}Accuracy \\ perturbed (\%)\end{tabular} & \begin{tabular}[c]{@{}c@{}}Accuracy un-\\ perturbed (\%)\end{tabular} & \begin{tabular}[c]{@{}c@{}}Cluster prediction \& KNN \\ search time (on-chain) (s)\end{tabular} \\ \hline
500 & 91.2 & 100 & 10.58 \\ 
1000 & 92.4 & 100 & 19.72 \\ 
5000 & 90.6 & 100 & 142.13 \\ 
12500 & 90.8 & 100 & 295.38 \\ \hline
\end{tabular}%
}
\end{table}

Further, we evaluate the speed and accuracy of shard prediction and image retrieval as a function of our system's image corpus size for variants C-xxx and E-xxx. C-OOO stores the data and executes the lookup on-chain. In contrast, E-OOF/FOF emit the image data as events on the blockchain and performs image lookup, retrieval, and verification off-chain. Table \ref{tab:exp2a} for C-OOO shows that the on-chain retrieval speed drops significantly as corpus size increases, suggesting C-OOO is unfit for GenAI contexts with large amounts of data.  Table \ref{tab:exp2b} shows that E-FOF maintains high retrieval accuracy as corpus size increases, with an average retrieval speed of just over 4 ms for a corpus size of 1M images. Tables \ref{tab:exp2a} and \ref{tab:exp2b} were measured for 500 queries.   Further, we find that ingesting images for the system's initial setup takes an average of 683.2 ms per image in C-OOO, whereas E-FOF significantly improves speed requiring an average of only 81.5 ms per image.

We conclude that E-FOF exhibits scalability with corpus size and shard count, leading us to recommend variant E-FOF for the GenAI training opt-in/out task.

\begin{table}[h]
  \caption{Evaluating the recommended event-log storage variant (E-FOF) for accuracy and speed as corpus size increases,  showing good scalability. Shard count $k=25$.}
  \label{tab:exp2b}
\resizebox{\columnwidth}{!}{%
\begin{tabular}{cccc}
\hline
\begin{tabular}[c]{@{}c@{}}Corpus\\ (x10\textsuperscript{3})\end{tabular} & \begin{tabular}[c]{@{}c@{}}Accuracy \\ perturbed (\%)\end{tabular} & \begin{tabular}[c]{@{}c@{}}Accuracy un-\\ perturbed (\%)\end{tabular} & \begin{tabular}[c]{@{}c@{}}Cluster prediction \& KNN \\ search time (off-chain) (ms)\end{tabular} \\ \hline
100 & 91.6 & 100 & 0.58466 \\ 
250 & 91 & 100 & 1.50747 \\ 
500 & 91.2 & 100 & 2.53011 \\ 
1000 & 91.2 & 100 & 4.27562 \\ \hline
\end{tabular}%
}
\end{table}

\subsection{Evaluating Cost}
\label{sec:costs}

Transaction cost is a consideration in scaling DLT systems. C-OOO is significantly more costly than E-OOF/FOF. Ingesting images costs, on average, 0.9M gas/image for C-OOO and, in comparison, only 0.2M gas/image for E-xOx variants. Similarly, when adding an image,  a user would pay, on average, 19M gas/image for C-OOO but only 15M gas/image for E-xOx variants. Projecting the fingerprint embedding space onto a lower dimensional space using principal component analysis can further reduce these costs but does not alter the trend. The cost factor reinforces our design recommendation to use the DLT event log rather than in-contract storage for the key-value data.

\subsection{DECORAIT applied to Dreambooth}
\begin{figure*}[h]
    \centering
    \includegraphics[width=1\textwidth]{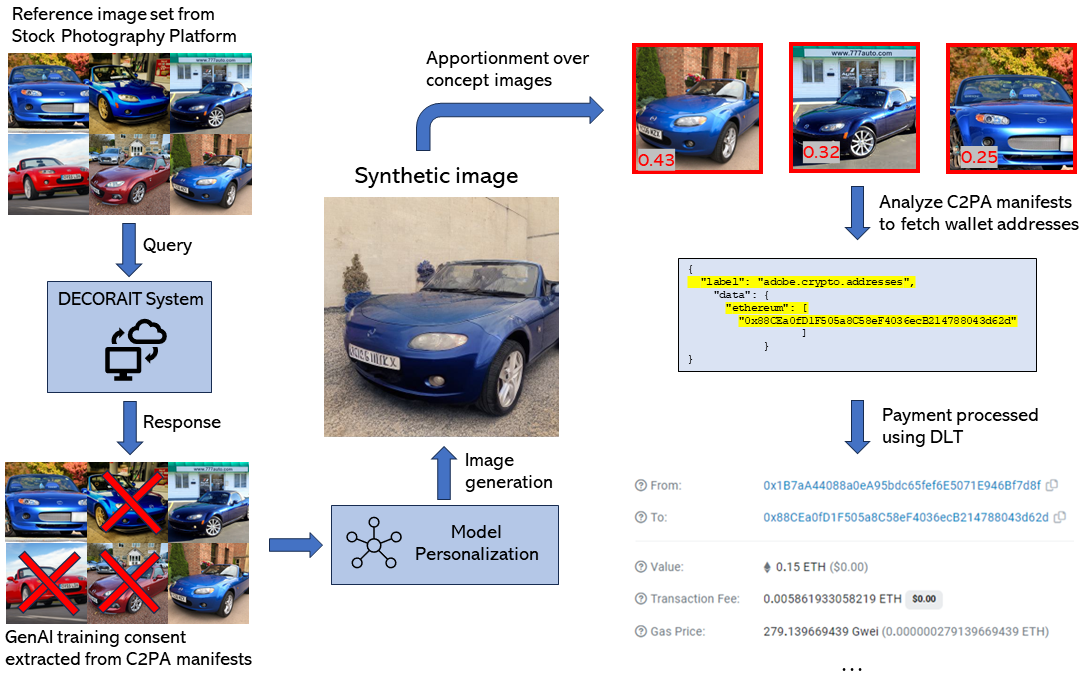}
    \caption{DECORAIT and Dreambooth pipeline including registry querying and model personalization flow. The Dreambooth model is specialized using the 3 opted-in images of a car and the proposed apportionment algorithm is applied across the image corpus. The red cross indicated images which have been opted-out according to the DECORAIT registry. The resulting apportionment conducted on the generated synthetic image from the experiment as described in Sec.\ref{eval_dreambooth} is shown. The DLT wallet addresses of the three authors of the images are identified using the accompanying C2PA manifests. Payment is then conducted automatically, securely, and transparently using DLT, and one transaction's confirmation is pictured.}
    \label{fig:apportionment}
\end{figure*}

\label{eval_dreambooth}

Using the recommended E-FOF variant of the system, we demonstrate DECORAIT in a real-world scenario by specializing a Stable Diffusion model using the Dreambooth \cite{dreambooth} method to synthesize renditions of a specific subject in new contexts. 

Initially, a set of concept images is purchased from a popular stock photography website, which can be viewed on the left side of Fig.\ref{fig:apportionment}. Unfortunately, their delivery is not accompanied by a C2PA manifest, therefore training consent cannot be immediately determined. The DECORAIT system is then queried to determine training consent across the set of concept images, by matching the images to their corresponding images within the registry. The assets within the registry are accompanied by C2PA manifests, which detail the author's choice of whether to allow GenAI training using that asset. The query to the DECORAIT registry resolves to several of our chosen images indicating that the creative has opted out of GenAI training. Fig.\ref{fig:teaser} pictures the effect differing training data can have on the resulting model and the synthetic media it is able to generate, especially when a subset of the chosen concept images has been opted-out of GenAI training. 

Once the model is trained using the opted-in images and following the Dreambooth method, we encode this provenance information within the manifest of both the resulting model and the generated synthetic image. An example provenance graph is pictured in Fig.\ref{fig:provenance} and we follow the same structure in this example. The "ingredient" feature of the C2PA standard is leveraged in order to reference the resulting personalized model as the ingredient asset of the generated synthetic image. Within the personalized model's manifest, we encode as ingredients both the set of concept images it was trained on, as well as the base text-to-image Stable Diffusion model which was fine-tuned in order to create the personalized model. The base model may include within its manifest an archive detailing its entire training corpus of ingredient images.

Further, we apply the apportionment algorithm in order to reward the contributing creatives. The process starts from the C2PA manifest of the synthetic image, tracing the provenance graph in order to identify the personalized model which created it and ultimately its training images. Then, the apportionment accumulates prediction scores using the fingerprinter and second stage classifier model for each concept image the model was specialized on. The wallet addresses belonging to the creatives who authored the training images are identified by analyzing the C2PA manifests of those images. Lastly, payments are processed for each contributing creative, with currency sent directly to their wallet address through DLT, as pictured on the right side of Fig.\ref{fig:apportionment}. The transaction confirmation is also pictured.

Thus, we have demonstrated an end-to-end pipeline which included ethically building a dataset of assets which have been opted-in for GenAI training, successfully avoiding copyright infringement, personalizing a generative diffusion model, as well as analyzing the resulting synthetic media and running our proposed apportioning algorithm in order to recognize and reward the contributing creatives, enabling near-instant processing of royalty-like payments using DLT.

\section{Conclusion}

We presented an end-to-end system through which content creators may assert their right to opt in or out of GenAI training, as well as receive reward for their contributions. We investigated the feasibility of a decentralized opt-in/out registry for GenAI using DLT, reaching recommendations that 1) event-log storage is appropriate; 2) on-chain shard prediction is appropriate for ingest but not for the query. We propose variant E-FOF as the most scalable solution, achieving 100\% accuracy on non-augmented queries and 91.2\% accuracy in the presence of augmentations, with query speed up to 4 ms for a corpus of 1M images.

DECORAIT employs the distributed ledger (DLT) as a trustless registry and source of truth. The bulk of the computationally expensive operations are conducted off-chain. We proposed a fingerprinting-based content similarity score for image attribution and credit apportionment over the attribution corpus in the case of synthetic media, with payments securely processed for the contributing creatives using DLT. The system leverages the C2PA standard to track content provenance, specify GenAI training consent and store the creator's DLT wallet addresses.  We demonstrated the DECORAIT system as part of a Dreambooth GenAI model personalization pipeline, demonstrating our proposed method for recovering synthetic media provenance and apportioning credit.  DECORAIT thus enables contributing creatives to receive recognition and reward when their content is used in GenAI training. 

Future work could incorporate the DECORAIT registry within popular GenAI data loaders and ship the apportioning flow as a library in order to drive adoption. Most notably, future efforts should focus on investigating the socio-technical drivers and challenges our system may face when deployed in the wild. Further consideration is required for its development and implementation within a sustainable business model. Equally critical is the necessity for establishing comprehensive policies within the legal and regulatory space addressing digital rights and data sourcing for training GenAI models. These questions are likely to remain open for some time, however, ensuring the consensual use of digital assets and fair reward to contributors within the GenAI training pipeline is both a timely and urgent matter. We believe the proposed DECORAIT system is a promising first step towards a decentralized, end-to-end solution to the problem.

\begin{acks}
DECORAIT was supported in part by DECaDE under EPSRC Grant EP/T022485/1.
\end{acks}

\bibliographystyle{ACM-Reference-Format}
\bibliography{decorait}

\end{document}